\newcommand{\wb}{\omega_{\mathrm{b}}}
\newcommand{\wo}{\omega_0}
\newcommand{\ii}{\mathrm{i}}

\newcommand{\ee}{\varepsilon}

\newcommand{\doublefig}{\textwidth}
\newcommand{\singlefig}{0.75\textwidth}

\def\fracc#1#2{\displaystyle \frac{#1}{#2}}
\documentclass{iopart}
\usepackage{iopams}
\usepackage{graphicx}

\renewcommand{\br}{\mathbf{r}}

\newcommand{\bt}{\mathbf{t}}
\newcommand{\bp}{\mathbf{p}}

\begin{document}

\title[Breathers in curved alpha--helix]
{Stationary and moving breathers in a simplified model of curved alpha--helix proteins}

\author{JFR Archilla \dag, Yu B Gaididei \ddag,  PL Christiansen \S \mbox{} and J Cuevas \dag}

\address{\dag
 Nonlinear Physics Group of the University of Sevilla.
 Dep. F\'{\i}sica Aplicada I, ETSI Inform\'atica.
 Avda. Reina Mercedes s/n, 41012 Sevilla, Spain.}

\address{\ddag Bogolyubov Institute for Theoretical Physics. 03143 Kiev,
Ukraine.}

\address{\S Informatics and Mathematical  Modelling.\\
The Technical University of Denmark, DK-2800 Lyngby, Denmark.}

\ead{\mailto{archilla@us.es}}

\begin{abstract}
The existence, stability and movability of breathers in a model
for alpha-helix proteins is studied. This model basically consists
a chain of dipole moments parallel to it. The existence of
localized linear modes brings about that the system has a
characteristic frequency, which depends on the curvature of the
chain. Hard breathers are stable, while soft ones experiment
subharmonic instabilities that preserve, however the localization.
Moving breathers can travel across the bending point for small
curvature and are reflected when it is increased. No trapping of
breathers takes place.
\end{abstract}

\submitto{\JPA 7 June 2002}

\section{Introduction}
\label{sec:introduction} In the last years quite a great deal of
research is being done on localization due to the interplay of
nonlinearity and geometry, either with the Nonlinear Schr\"odinger
Equation \cite{GMC00,CGM01}, FPU models \cite{RSIT01,IST02,TIS02},
Klein--Gordon models \cite{ACMG01,ACG01} or in DNA models
\cite{TP96b,FCP97}. The objective is to understand the role of the
bending points in biomolecules: whether localized excitations can
travel across them or not and whether there are points where
energy is stored and play a biological function. The description
of the biological systems is given by variables, that represent
internal or external degrees of freedom, which oscillate with time
and are coupled by different potentials. A change of geometry can
be felt by the system by different physical mechanisms, modelled
correspondingly: interaction between nearest and next neighbours
\cite{IST02,TIS02}; potentials that depend on the angles
\cite{GMC00} or long--range interaction due to the dipole--dipole
coupling \cite{CGM01,ACMG01,ACG01}.

Simplifying, the effect of a curved chain can be described easily.
For stationary excitations the zone where the chain of oscillators
is bent is inhomogeneous, bringing about the existence of
localized linear modes which compete strongly with the nonlinear
localized modes. For high coupling the linear localization
predominates, for low coupling the nonlinear localization is the
important one. The transition from one regime to the other can be
continuous or discontinuous as the parameters are changed or the
curvature increases \cite{ACG01}. Moving excitations, have to
travel across this inhomogeneous zone. The excitation within the
bending zone needs a different energy, which can be larger than in
the straight system, therefore acting as a barrier. The excitation
can be reflected, transmitted or trapped \cite{CPAR02}.

The alpha--helix protein is another molecule for which is interesting to
investigate the role of the bending in the existence, shape, properties
and transport of localized excitations. The peptide groups have a dipole
moment parallel to the chain and the Amide-I excitations interact among
them through acoustic phonons. The dipole-dipole interaction to be
described in detail below makes the system sensitive to the shape of the
molecule. Apart from the interest of the biophysical problem in itself,
it has a relevant feature from a more theoretical point of view: it is
only necessary to take into account the nearest neighbours in order that
the system can feel the bending of the chain. To our knowledge the
effect of the curvature in such a system with dipoles parallel to the
chain have not yet been described.

There are different excitations that can be considered, as solitons,
envelope solitons and discrete breathers. Here we focus our research on
the latter. As it is well known, discrete breathers are very localized
oscillations that appear as a consequence of nonlinearity and
discreteness \cite{MA94,MA96, A97,FW98}. They are, therefore, specially
relevant in biomolecules when considering excitations that involve only
a few units, far from the continuous limits. Although a relatively new
field, stationary breathers are now well understood and research is
presently focusing on the possible physical and biological consequences
of their existence and in future technological applications, specially
with Josephson-junctions \cite{BAU00a,TMO99}. Moving breathers is not
such a mature field: there are techniques to obtain them but many
questions remain unanswered \cite{CAT96,AC98}. Perhaps the most
important one is the reason why they exist in some systems and not in
others. New developments are given in \cite{MS02}.

This study intends to complete previous works
\cite{ACMG01,ACG01,CPAR02} on the properties of localization and
transmission of energy in bent chains, described by Klein--Gordon
models. The models in the references cited are inspired in DNA,
with three main properties:
 \begin{enumerate}
 \item There is long--range interaction between dipole moments.
 \item The dipoles moments are perpendicular to the chain and to
 the plane of curvature.
 \item
The effect of the curvature enters the model as a shortening of
the distances between dipole moments.
\end{enumerate}

The simplified alpha--helix model studied here, apart from its
different physical origin, has the following differences:
\begin{enumerate}
\item The only interactions considered are nearest--neighbour.
Note that the DNA models cited, without long--range interaction
would not feel at all the shape of the molecule.
\item
The dipole moments are parallel to chain and to the plane of
curvature
\item The bending of the chain appear in the model as a change in
the angles between dipole moments, and, therefore in the
interaction energy.
\end{enumerate}

In spite of these differences, both models present very similar
behaviour both for predominant stronger attractive or repulsive
interaction. This suggest that they are generic on bent
Klein--Gordon systems.

However, this paper introduces a new point of view. Previous works
misinterpreted the mathematical property of annihilation of
breathers as the curvature is increased. What happens is that the
bending forces the breather at the bending point to choose some
frequency, which depends weakly on its energy.

This phenomenon, could be, in principle, tested experimentally.
For example, the mean curvature of DNA depends on the solvents
concentration, and the absorbtion of radiation by breathers at the
bending points would appear at some characteristics frequencies,
that change with the solvents concentration. The presence of some
harmonics of these frequencies would differentiate breathers from
linear localized modes. Certainly, many technical problems should
be expected, but we are planning to perform this study with some
experimental groups. Apparently, the more adequate molecule might
be RNA due to its short persistence length.

The short section~\ref{sec:moving} deals with moving breathers.
Its first objective is to check if in this model, breathers can
move and are reflected by or transmitted through the bending
point. The second is to confirm the {\em trapping hypothesis}
introduced in Ref.~\cite{CPAR02b} for a DNA model with an
impurity. The role of an impurity is similar to a bending point in
bringing about localized linear modes, which compete with the
(nonlinear) breather localization. This hypothesis  consist in the
nonexistence of trapping for moving breathers when there exists a
linear localized mode with different tail profile, i.e.,
neighbouring sites in phase or with opposite phase (and,
therefore, different frequency). There exist no mathematical
proof, but it is also confirmed in this work in spite of the
differences stated above.

\section{Model description}
\label{sec:model}
We describe the position of each amino acid by $\br_n$, $n$ being an
index. The distance between amino acids is supposed to be a constant
$a$, i.e. , $a=|\br_{n+1}-\br_n|$. The direction of the dipole moments
are given by unit vectors
$\bt_n=\fracc{(\br_{n+1}-\br_n)}{|\br_{n+1}-\br_n|}$ and the moments
themselves are given by $\bp_n=p_n\,\bt_n$.

The Hamiltonian of the system, for which details are given in
~\ref{sec:Adeduction}, can be written as
\begin{eqnarray}
\label{eq:mainhamiltonian}
 \fl
 H=\sum_n \ \fracc{1}{2} {\dot
u}^2_n +\fracc{1}{2}\omega_0^2 u_n^2 +\Psi(u_n) +
  \fracc{1}{2} \ee (u_{n+1}-u_{n})^2  +
\mu (\bt_{n+1}-\bt_n)^2   u_n u_{n+1}\,,
\end{eqnarray}
with $\omega_0=1$,  and the on site potential is given by
$V(u_n)= \frac{1}{2} \wo^2 u_n^2 +\Psi(u_n)$ ,
$\Psi(u_n)$ being its nonlinear part.

The corresponding dynamical equations are
\begin{eqnarray}
\label{eq:dynamic1}
 \fl
 {\ddot u}_n+ \omega_0^2 u_n +\Psi '(u_n) +
\ee (2\,u_n -u_{n+1}-u_{n-1})  \nonumber \\
 \lo +\;\mu
((\bt_{n+1}-\bt_n)^2 u_{n+1}+  (\bt_{n-1}-\bt_n)^2 u_{n-1})=0 \,,
\end{eqnarray}
or, in a simplified notation
\begin{eqnarray}
\label{eq:maindynamic}
 \fl
  f_n(u)\equiv {\ddot u}_n+\omega_0^2 u_n +\Psi '(u_n) +
  \ee \sum_m C_{n,m} u_m + \mu \sum_m
 J_{n,m} u_m=0 \,,\;
\end{eqnarray}
with the obvious definition of the coupling matrices $C$ and $J$.
Note that $C$ does not depend on the parameters while $J$ depends
only on the curvature $\kappa$

Certainly, many different shapes can be considered. Here we focus
as in previous papers on a parabola in a two--dimensional space.
The reason is that it is the simplest form to describe a bending
point and it is an approximation of any curve in the neighbourhood
of it. Other shapes considered in the literature like the hairpin
geometry, although interesting in themselves are not so adequate
for our purposes, because, for example, the latter is composed by
three homogeneous regions (except if considering long--range
interaction). Therefore, the vectors $\br_n$ have components
$(x_n,y_n)$ with $y_n=\frac{1}{2}\kappa x_n^2$, $\kappa$ being the
curvature.

\begin{figure}
\begin{center}
\includegraphics[width=\singlefig]{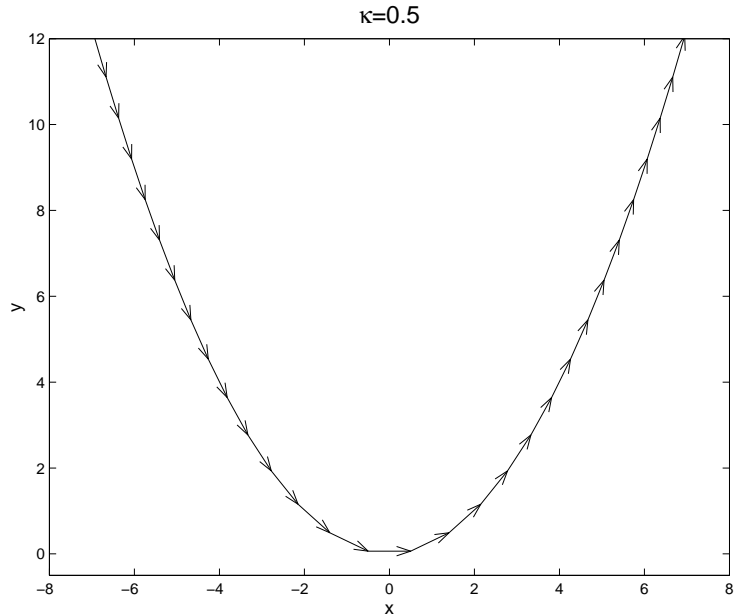} 
\end{center}
\caption{Sketch of the model for curvature $\kappa=0.5$.}
\label{fig:sketch}
\end{figure}

A sketch of the model is shown if figure~\ref{fig:sketch}

In spite of having many dimensionless variables, there are still a
few parameters in our problem: $\ee$, which will be called the
stacking coupling parameter, $\mu$, the dipole coupling parameter,
the curvature $\kappa$ and the breather frequency, which we
represent by $\wb$. The distance of $\wb$ from the phonon band is a
measure of the degree of nonlinearity of the excitation. Generally
speaking, the more nonlinear the excitation, the narrower it
becomes. To explore thoroughly the parameter space is a daunting
task but to obtain a general picture is a much more affordable one
and this is what we intend here.

\section{Linear modes}
\label{sec:linear}

\begin{figure}
\begin{center}
\includegraphics[width=\singlefig]{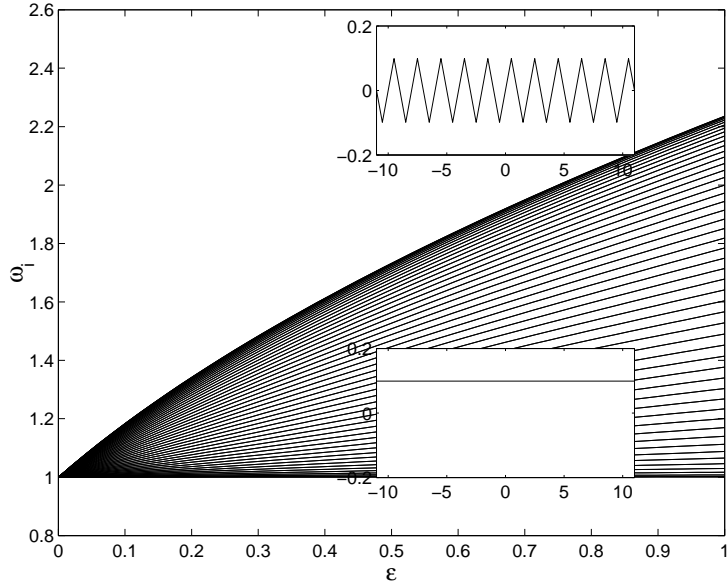}
\end{center}
\caption{
Dependence of the linear spectrum of the system on the stacking
coupling parameter $\ee$ for a straight chain. The insets show the shape
of the linear modes with highest and lowest frequency.}
\label{fig:linearee}
\end{figure}

In bent systems there exist linear modes that are localized around
the bending point. The first step is therefore to obtain the
dependence of the linear spectrum on the parameters and identify
the most important ones. The linear modes are the solutions of
equation~\ref{eq:maindynamic} with $\Psi=0$. We first obtain the
dependence on the stacking parameter $\ee$ without dipole coupling
or equivalently for a straight chain, i.e., either $\mu=0$ or
$(\bt_{n+1}-\bt_n)=0, \;\forall n$ in
Equation~\ref{eq:maindynamic}. Figure~\ref{fig:linearee} shows
this spectrum dependence.

This a well known spectrum, which we include here for comparison.
Let us comment three facts: first, the spectrum is continuous;
second, the mode with highest frequency consist of all oscillators
vibrating with opposite phase with respect to the nearest
neighbours; third, the mode with lowest frequency consist of all
oscillators vibrating in phase. This modes will hereafter be
denoted the top and the bottom mode, respectively.

\begin{figure}
\begin{center}
\includegraphics[width=\singlefig]{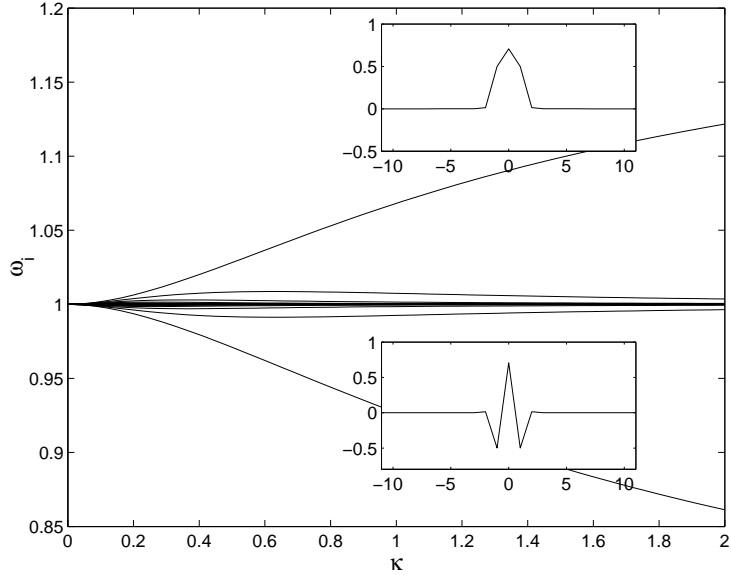}
\end{center}
\caption{Dependence of the linear spectrum of the system on the curvature
$\kappa$ with stacking coupling parameter $\ee=0$ and dipole
coupling parameter $\mu=0.2$. The insets show the bell and zig-zag
shapes of the linear modes with highest and lowest frequency,
respectively.}
\label{fig:linearkappa}
\end{figure}

When the stacking coupling $\ee=0$ we have to choose either to
represent the variation of the linear spectrum as a function of
the parameter $\mu$ or to the curvature $\kappa$, the other
parameters fixed at non--zero values. It turns out that both
possibilities give similar results. Figure~\ref{fig:linearkappa}
shows the dependence of the linear spectrum on $\kappa$. There are
three key features: first, there are localized modes that separate
from the continuous spectrum; second, the top and bottom modes are
localized within a radius of about a few units around the bending
point and they become more localized as $\kappa$ increases; third,
the top mode has all the oscillators in phase, which we will
describe as being bell shaped, while the bottom mode has
neighbouring oscillators with opposite phase (zig--zag shaped).
This change of profile is due to a change of the predominant
coupling interaction when $\kappa$ (or $\mu$) is increased. Around
the bending point the sign in front of the variable $u_m$ in the
coupling terms of equation~(\ref{eq:maindynamic}) changes from
negative, i.e, attractive interaction to positive, i.e., repulsive
interaction. For homogeneous repulsive interaction the linear
spectrum would be as in figure~\ref{fig:linearee}, but the
frequencies spreading downwards and the zig--zag mode at the
bottom.

The localization does not depend on the number of sites in the
system. If the dipole coupling parameter $\mu$ is changed at
constant curvature $\kappa$, the spectrum is very similar with a
slightly larger spread of the frequencies except for the top and
bottom modes.

\begin{figure}
\begin{center}
\includegraphics[width=\singlefig]{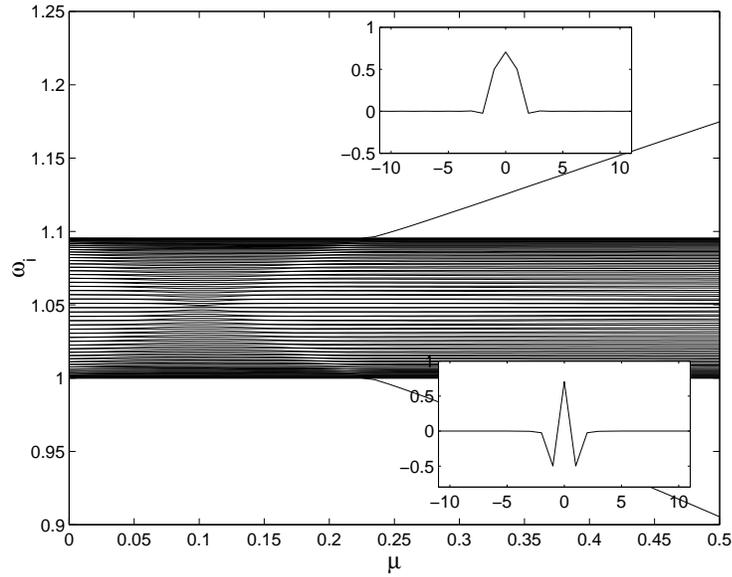}
\end{center}
\caption{Dependence of the linear spectrum of the system on the dipole coupling
parameter $\mu$ for constant stacking parameter $\ee=0.05$ and curvature
$\kappa=1$. The insets show the shape of the linear modes with highest
and lowest frequency, respectively.}
\label{fig:mixed}
\end{figure}

To conclude the description of the linear spectra we represent in
figure~\ref{fig:mixed} the dependence of the spectrum on the dipole
parameter $\mu$ for a non--zero, constant value of the stacking
parameter $\ee=0.05$ and constant curvature $\kappa=1$. It can be seen
that for some value of $\mu$ the same localized modes separate from the
continuous spectrum. Note that while studying breathers these modes are
going to be the ones competing with nonlinear excitations and be
predominant for high enough coupling. The frequency $\wb$ of the
breathers have to be outside the continuous band or it will not be
possible to obtain the breathers, which in physical terms means that the
linear localized modes will resonate with the breather frequency and the
energy will spread along the chain. If the on--site potential is hard,
the breather frequency $\wb$ will be above the phonon band, $\wb>\wo$,
therefore competing with the bell--shaped, top, linear mode. In the
opposite case, $\wb<\omega_0$, the competing mode will be the zig--zag
shaped, bottom one. This will to be described in detail in the next
section.

\section{Hard breathers}
\label{sec:hard}

\begin{figure}
\begin{center}
\includegraphics[width=\singlefig]{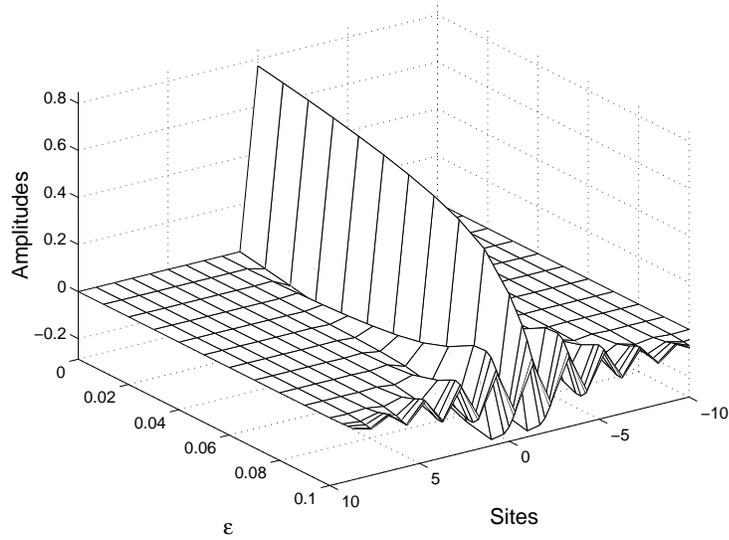}
\end{center}
\caption{
Dependence of the breather profile with hard, on--site potential for the
straight chain on the stacking parameter $\ee$. Frequency $\wb=1.2$.}
\label{fig:brstacking}
\end{figure}

To obtain breathers we need to choose the value of their
frequency, which for a hard potential (here
$V(u_n)=\frac{1}{2}\wo^2u_n^2+1/4\,u_n^4$\;) will be above
$\omega_0=1$. Let us choose $\wb=1.2$, i.e., a frequency not too
far from the phonon band but outside of it. The properties of the
breathers with stacking coupling are well known, but we include
them here for comparison. Figure~\ref{fig:brstacking} shows the
profile of the breather while the parameter $\ee$ is changed. The
path continuation using the Newton method finishes as the phonon
band expands and collides with the breather frequency. The
breather profile is a zig-zag one, which can also be easily
deduced by tail analysis.

\begin{figure}
\begin{center}
\includegraphics[width=\singlefig]{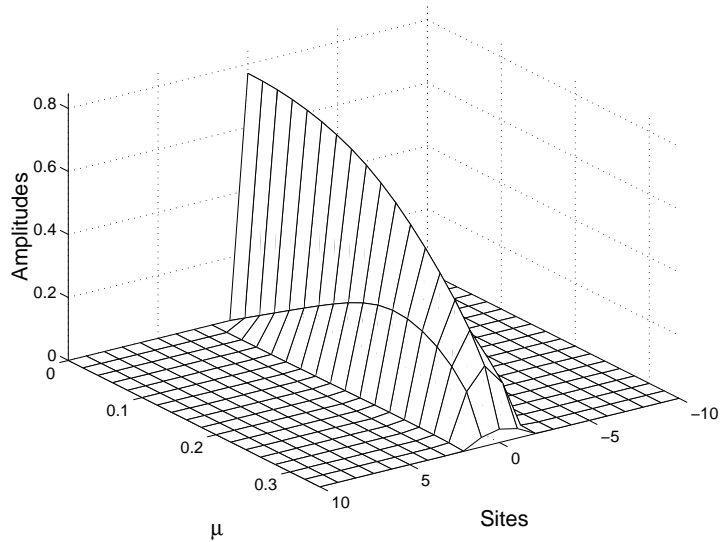}
\end{center}
\caption{
Dependence of the breather profile with hard on--site potential for
a curved chain with curvature $\kappa=2$ on the dipole coupling
parameter $\mu$ with constant frequency $\wb=1.2$. $\ee=0$}
\label{fig:brdipole}
\end{figure}

We obtain an analogous picture if the continuation is done with respect
to the dipole coupling parameter $\mu$, for constant curvature, until a
bifurcation point.
 Now the breather gets the
bell profile shown in figure~\ref{fig:brdipole}.

The fact that the breather amplitudes, i.e, $\{|u_n(0)|\}=\{
\max_{\forall t}(|u_n(t)|)\}$, tend to zero is misleading. We are
in presence of an annihilation bifurcation \cite{CAGR02} , i.e.,
the Jacobian of the dynamical equation (\ref{eq:maindynamic}) with
respect to $u=(u_1,u_2,\dots,u_n)$ has an eigenvalue that tends to
zero. This eigenvalue does not correspond to another breather in
the neighbourhood of the parameter space but to the upper (for a
hard potential) breather band \cite{A97} of the same breather. The
physical meaning is that there are no localized excitation with
the chosen frequency, except in the trivial case where the
breather has zero amplitude, i.e., it is in the linear regime. We
can obtain a complementary picture if the continuation is
performed while keeping constant another characteristic of the
breather. In \cite{AMM99} this is done with constant action, here,
using a similar technique, we have chosen to keep the energy
constant, which is physically meaningful.

By fixing the energy we restrict ourselves to the  situation which
is realized in single-molecule experiments.  More biologically
meaningful would be to keep constant  not the energy but the
temperature.  This is however beyond the scope of our  paper
because to carry out  this type of investigation it is necessary
to introduce into equations of motion stochastic forces and solve
corresponding Langevin equations.

\begin{figure}
\begin{center}
\includegraphics[width=\singlefig]{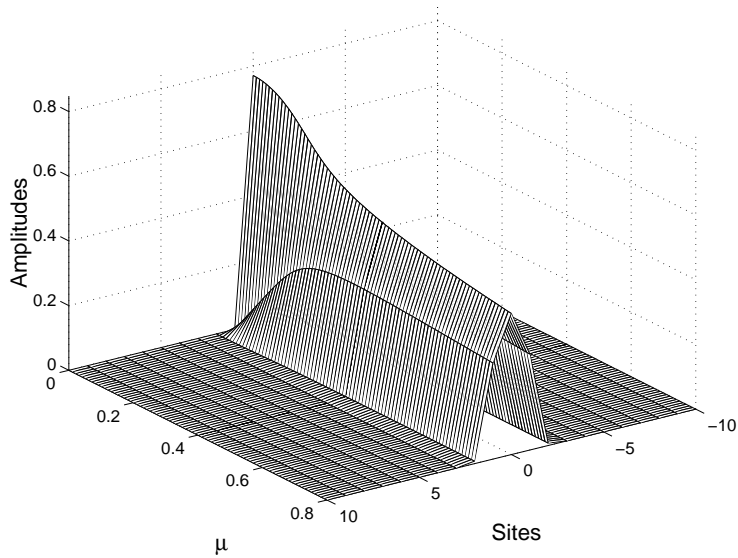}
\end{center}
\caption{
Dependence of the breather profile with hard on--site potential for a
curved chain with curvature $\kappa=2$ on the dipole coupling parameter
$\mu$ for constant energy, $E=0.3855$. The frequency changes from 1.2 to
1.46. $\ee=0$.}
\label{fig:bredipolemu}
\end{figure}

The dependence of the breather profile on the parameter $\mu$ at
constant energy is plotted in figure~\ref{fig:bredipolemu}. It can
be seen that allowing the breather the freedom of choosing its
frequency the continuation is possible for much larger values of
$\mu$, or, in physical terms, there exist localized breathers
although not at any frequency. Note that in this situation the
phonon band is very narrow, and what we have is a nonlinear
version of the linear mode described previously, whose frequency
is close to the breather's one although not identical.

\begin{figure}
\begin{center}
\includegraphics[width=\doublefig]{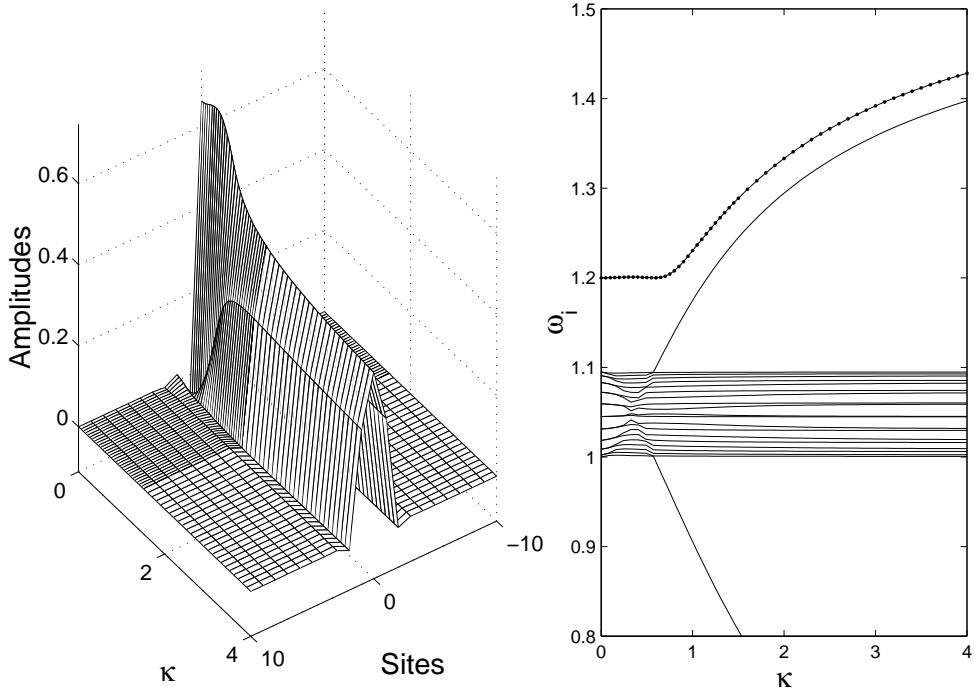}
\end{center}
\caption{Left: Dependence of the breather
profile with hard on--site potential for a curved chain on the
curvature at constant energy. Right: Plot of the breather frequency
(dots) and the linear spectrum with respect to the curvature.
$E=0.3064$, $\ee=0.05$, $\mu=0.5$. }
\label{fig:brkappa}
\end{figure}

Although the cases described above are interesting to understand
the phenomena, the situation with more physical interest is the
process of curving the chain. Suppose that $\ee\neq 0$ and $\mu
\neq 0$ and the breather has its frequency above the phonon band.
When the chain is curved, the dependency of the breather with
respect to the curvature is similar as with respect to $\mu$. If
we choose to maintain constant the frequency the plot of the
breather evolution is very similar to figure~\ref{fig:brdipole},
except for the fact that at the zero curvature the breather would
have a zig--zag profile which progressively changes to the bell
profile for large enough curvature. Perhaps more interesting is
choice of keeping the energy constant, as it is a conserved
quantity. The dependence of the breather amplitudes and frequency
are shown in figure~\ref{fig:brkappa}. In the figure to the right
we can see how the linear mode appears and the frequency of the
top mode increases until almost colliding with the breather
frequency. The breather with frequency $\wb=1.2$ no longer exists
while there exists a breather which is the nonlinear analogue of
the top linear mode and its frequency increases as the chain is
curved.

The linear mode is a solution of equation~(\ref{eq:maindynamic}),
with $\Psi=0$, or, equivalently, with almost zero amplitudes for
$\Psi\neq 0$. What we have called its nonlinear analogue, is
solution of the same equation with $\Psi\neq 0$ and  has nonzero
amplitudes, has a similar profile, as can it is shown in
figure~\ref{fig:brkappa}.   It has more than a single harmonic in
its linear spectrum. These extra harmonics are significatively
larger for breathers with soft on--site potentials, as studied in
the next section. In the case studied here, as  the linear mode is
localized, and, therefore, its frequency is isolated from the
phonon band, the nonlinear analogue can be obtained by
substituting $\Psi\rightarrow s \Psi$ in
equation~(\ref{eq:maindynamic}), and changing continuously $s$
from zero to $1$ at constant energy.

The observation of figure~\ref{fig:brkappa} shows that for
frequencies above the top of the phonon band, i.e., slightly below
$1.1$ for the parameters in the figure, it is only possible to
continue the breather with the curvature at constant frequency
until it approaches the top mode frequency curve. Thereafter, it
is possible to continue it at constant energy, and its frequency
follows slightly above the top mode.

If the chain is curved around the neighbouring site where the
initial excitation is located, it will switch to the bending
point, for even small curvature, when the frequency of the top
linear mode approaches to its frequency. This happens at
$\kappa=0.7$ for the values corresponding to
figure~\ref{fig:brkappa}. Similar behaviour has been found in
reference~\cite{ACMG01}.

Let us mention that all the breathers described here are linearly stable, i.e.
all their Floquet eigenvalues have modulus $1$. The hard breathers in this system
are always stable. The properties of the soft ones are very different.

\section{Soft breathers}
\label{sec:soft}

We have included the previous section for completeness, but,
actually, in biological molecules, chemical bonds are thought to
be better described by soft potentials. The most commonly used are
the quartic soft, $V(u_n)=\frac{1}{2}\wo^2u_n^2-\frac{1}{4}u_n^4$,
the cubic potential,
$V(u_n)=\frac{1}{2}\wo^2u_n^2-\frac{1}{3}u_n^3$, the
Lennard--Jones potential, $V(u_n)=\sigma(1/r^6-1/r^{12})$, and the
Morse potential $V(u_n)=D(\exp(-b\,u_n)-1)^2$. The last two are
known to have several nice characteristics: a) they are
asymmetric, with a hard part, that describes the strong repulsion
when two atoms or molecules approach, and a soft part that becomes
flat, reflecting the weakening of the bond when the molecules get
separated and eventually unbonded; b) they have moving breathers
with stacking coupling potentials; c) compared with another
frequently used soft potential, the cubic one, it does not have an
infinite well, which has no physical interpretation and can
produce anomalous results when simulations are done. We have
chosen for presenting our results the Morse potential because it
is mathematically simpler, but the Lennard--Jones potential
provides similar results.

For the straight chain and the normalized Morse potential $D=1/2$ and
$b=1$, which gives the same frequency $\wo=1$ for small oscillations,
and a representative frequency of the nonlinear excitations $\wb=0.8$ ,
the breathers are bell-shaped, their amplitudes grow with the coupling
parameter and they become unstable for $\ee=0.12$. The nonlinear mode
that produces the instability is an spatially asymmetric one. This will
be of special interest in the section of moving breathers.

For stationary breathers, we are interested in values of the
parameters where the dipole interaction is significant and the
nonlinearity is weak. In this situation, typical parameters can be
$\ee=0.05$, $\mu=0.5$. For these parameters the frequencies of the
phonon band, i.e., the continuous part of the linear spectrum,
spread from $1$ to $1.1$. A rough measure of the nonlinearity of
the excitation is the distance of its frequency from the phonon
band, which suggest a frequency around $0.8$ or $0.9$.

If we start with the straight chain and a breather centered at the
bending point and we increase the curvature at constant frequency, the
same difficulties as with the hard breathers arise: the amplitudes of
the breathers tend to zero when its frequency approaches the bottom
linear mode and the continuation becomes impossible. We must emphasize
this point: in the breather literature path continuation is usually done
at constant frequency, which is a natural consequence of the theorems of
existence \cite{MA94} from the anticontinuous limit. At the
anticontinuous limit there exists freedom to choose the frequency of the
isolated oscillators and therefore at low coupling too. This is not the
case if the system is in presence of strongly localized modes: there are
no localized excitations at every frequency. This characteristic opens
the possibility of spectroscopic analysis to check the existence of
breathers in biomolecules and eventually to measure curvatures.

\begin{figure}
\begin{center}
\includegraphics[width=\doublefig]{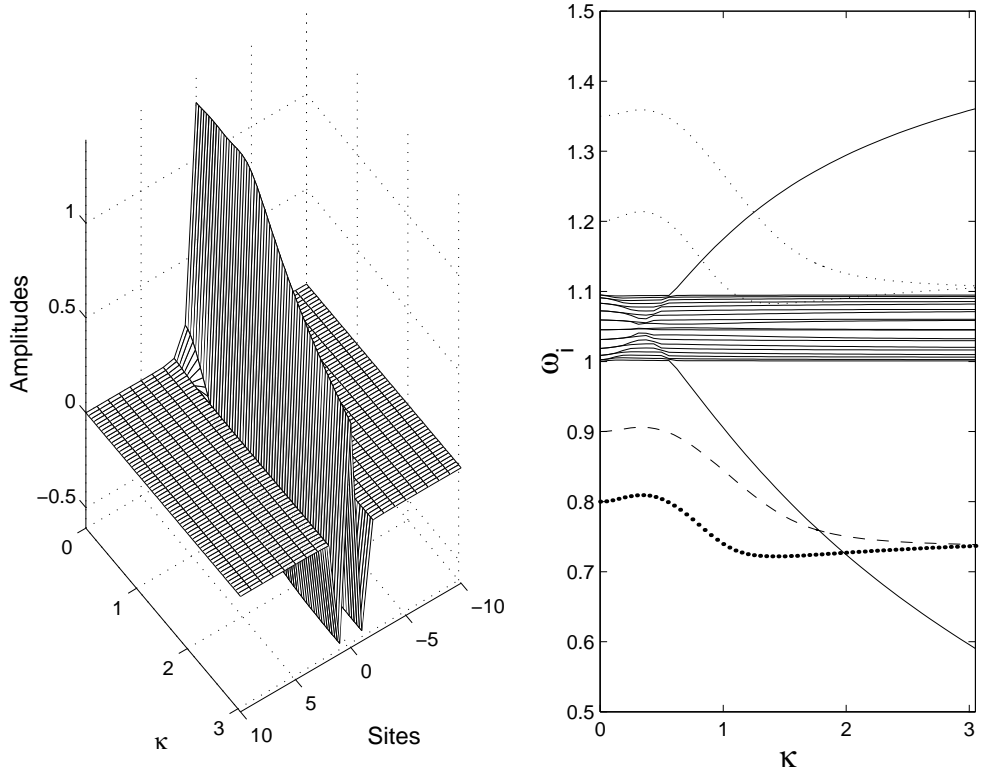}
\end{center}
\caption{Left:
Dependence of the breather profile with soft, on--site potential
for a curved chain on the curvature at constant energy E=0.35.
Right: Plot of the breather frequencies for two energies, E=0.35
(big dots) and E=0.20 (dashed), and 1.5 times the breather
frequencies (small dots) and the linear spectrum (continuous lines)
with respect to the curvature. Parameters: $\ee=0.05$, $\mu=0.5$. }
\label{fig:softbr}
\end{figure}

From the mathematical point of view, if the frequency is not fixed, we
need to fix another quantity, and the chosen one, as previously, is the
energy. Thus, the variation of the breather with the curvature mimics an
adiabatic process of bending. Figure~\ref{fig:softbr}--left, shows the
dependence of the breather profile with $\wb=0.8$ at $\kappa=0$ on the
curvature, changing its shape from a bell profile to a zig--zag one,
until its becomes almost invariable. While this process takes place, its
frequency changes, which is shown in figure~\ref{fig:softbr}--right. The
breather becomes unstable at $\kappa \approx 0.8$. The frequencies of
another breather with lower energy and higher frequency is also
represented in this figure. The higher the frequency the larger becomes
the curvature, for which instability is produced.

The exact stability analysis has been done calculating the Floquet
eigenvalues, but we prefer to plot here the breather frequency,
the linear spectrum and $1.5$ times the breather frequency. This
gives a much more intuitive physical picture.

The reason for plotting this frequency is the following. When the
curvature is increased, the frequencies $\{\omega_i\}$ of the
internal modes, i.e., the small perturbation of the breather, are
very similar to the ones shown in figure~\ref{fig:mixed}, as
commented in section~\ref{sec:linear}. The biggest difference is
that the bottom mode (with soft on--site potential, and the top
one for hard on--site potential) is not there, because this mode
is substituted by the breather itself. The corresponding Floquet
multipliers are $\{\lambda_i=\exp(\pm \,2\,\pi\,\omega_i/\wb)\}$,
$\{\theta_i=\pm \,2\,\pi\,\omega_i/\wb\}$  called the Floquet
arguments. For a instability to be produced, a collision of two
multipliers has two take place with the condition that they have
the same Krein signature (See reference~\cite{A97} for details on
this subject). The positive (negative) Floquet arguments have all
of them the same positive (negative) signature and cannot
therefore bring about instabilities until they collide at
($1+0\,\ii$), named harmonic instabilities, or ($-1+0\,\ii$),
subharmonic instabilities, or between them, named oscillatory
instabilities or Krein crunches. The first one happens when
$\theta_i=2\,\pi\,n$, $n\in\mathbb{N}$, for some $i$, that is,
$\omega_i=n\,\wb$, or in other words a linear mode resonates with
the breather frequency or its harmonics. Figure~\ref{fig:softbr}
shows that this kind of bifurcation does not happen here.

The subharmonic instabilities can occur when
$\theta_i=\pi+2\,n\,\pi$,, for some $i$, i.e.,
$\omega_i=\wb/2+n\wb$. Within the range of parameters studied,
this can only occur when some $\omega_i$ equals $1.5\,\wb$.
Therefore, subharmonic instabilities  appear in our system, when
some frequency of the internal modes (very close to the linear
ones, as commented above), collide with the $1.5\,\wb$. This is
easily seen plotting simultaneously the $1.5\,\wb$ curve and the
linear frequencies.

Figure~\ref{fig:softbr}--right shows the reason of this
instability: the curve $1.5\,\wb$ intersects with the frequency
curve of the top linear mode, bringing about the instability.
Nevertheless, the latter is a localized mode, centered at the
bending point and the localization persists although the
excitation has now superimposed a perturbation with twice the
breather period. Within the range of parameters shown there are
also a  Krein crunches, related with perturbations of the
frequency, but also preserving the localization. This happens
after the Floquet multipliers of the phonon band have crossed at
($-1+0 \ii$), or, in other words, their frequency have become
larger than $1.5\,\wb$. The eigenvalues of the internal linear
modes cross with other and do small excursions outside the unit
circle. This is also evident, as the curve $1.5\,\wb$ enters the
phonon band. The same figure shows the frequency of a breather
with higher energy corresponding to a frequency $\wb=0.9$ at
$\kappa=0$. Note that both breathers tend to the same frequency as
the curvature increases and that breathers with energy slightly
below $E=0.35$ avoid the secondary  bifurcations, because the
$1.5\,\wb$ curve does not intersect the phonon band.

The apparent intersection of the breather frequency with the bottom
linear mode is misleading because the latter does not exist as a
perturbation of the breather, being this one its nonlinear
analogue. The breather is not, however, linear, as its first AC
harmonic is of the order of the DC one, and the second, about one
third of it.

\section{Moving breathers}
\label{sec:moving}

In this section we explain briefly the behaviour of a moving
breather in a curved region of the alpha--helix.

In order to move a breather, we apply the marginal mode method
\cite{CAT96}. It basically consists in adding a spatially
asymmetric mode (the pinning mode) to the breather velocity. As is
shown in \cite{CAGR02}, the stacking coupling of the breather must
be strong enough in order that it can  be moved.

\begin{figure}
\begin{center}
\includegraphics[width=\singlefig]{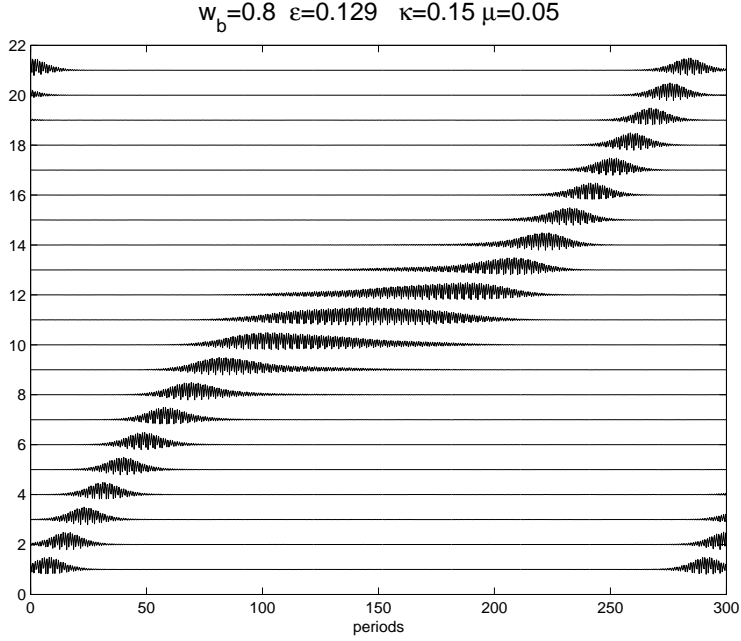}
\end{center}
\caption{A breather travelling across the bending point. A small
increase of the curvature will produce a reflection}
\label{fig:moving1}
\end{figure}

\begin{figure}
\begin{center}
\includegraphics[width=\singlefig]{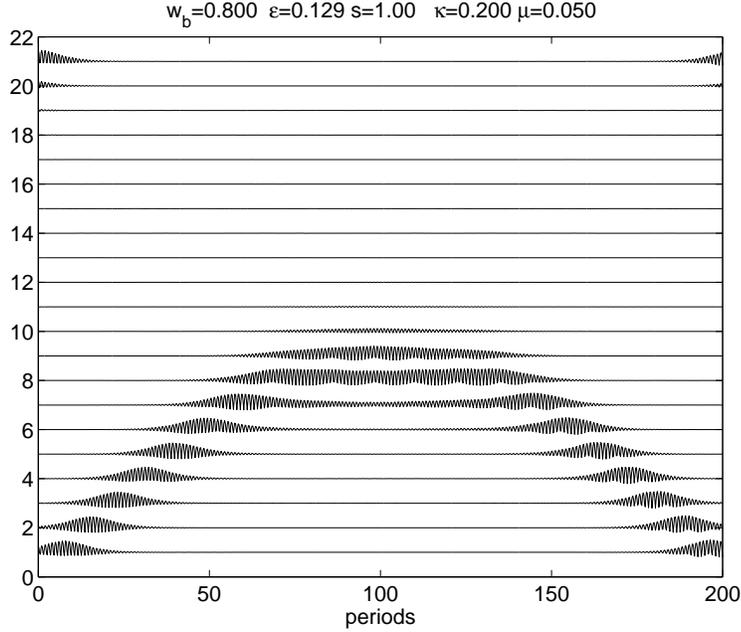}
\end{center}
\caption{A breather being reflected at the bending point.}
\label{fig:moving2}
\end{figure}

The scenario is similar to the observed in DNA chains
\cite{CPAR02} and biopolymers \cite{IST02}, i.e., for a fixed
value of the curvature, there exists a critical value of the
velocity below which the breather is reflected when it reaches the
bending point and, above which, the breather crosses through it.
Analogously, for a fixed value of the velocity, the breather
crosses the bending point as long as the curvature is smaller than
a critical value. As the curvature increases, the breather spends
more and more time at the bending point and eventually is
reflected. These two behaviours are shown in
figures~\ref{fig:moving1} and \ref{fig:moving2}. In other words,
the moving breather in a curved alpha--helix chain behaves as a
particle in a potential barrier.

In reference~\cite{CPAR02b} an hypothesis is introduced about the
existence of trapping in inhomogeneous Klein--Gordon lattices.
According to it, trapping of breathers does not occur when there
exists a linear localized mode with a profile (vibration pattern
or wave vector) different from the stationary breather one. When
this occurs, the moving breather behaves as a particle in a
potential barrier and is never trapped. As Figure \ref{fig:mixed}
shows, there is a localized mode (the top mode) of opposite
profile to the breather, and the case exposed in the reference
cited above is the same as in the curved alpha--helix.  The
simulations confirm that the trapping hypothesis \cite{CPAR02b}
holds in our system, and moving breathers cannot be trapped.

We still have not been able to find a counterexample, and this
hypothesis seems generic for Klein--Gordon systems. However,
 it has no mathematical proof and  should be checked as much as
 possible to achieve, at least, an inductive confirmation.

\section{Conclusions}

We have studied the existence and properties of stationary and
moving breathers in a model with dipole moments parallel to a
curved chain of oscillators. Its interest is twofold. On the one
hand it is a model for a channel of amino acids along an
alpha--helix protein, where the internal degrees of freedom are
the Amida--I vibrations. In this aspect the objective of our work
is to explore the role of the bendings as places where the energy
is stored and to investigate whether moving excitations can travel
across a bending point or not. On the other hand the theoretical
interest comes from it being a model that completes previous works
on bending chains. The specifics of the model are: a) the dipoles
are oriented along the chain instead of perpendicular to it; b) it
is sufficient to take in account nearest neighbours in order that
the system can feel the shape of the chain.

In previous models \cite{ACMG01,ACG01,CPAR02,CAGR02} the shape is
felt due to the change between the distances of the oscillators,
and, therefore, they need long--range interaction in inextensible
chains. In our model the shape is felt due to the change of the
angles between dipole moments. In this way, we expect to help to
complete the study of systems in curved chains and approach to a
generalized description.

The most important general fact, overlooked in the breather
literature to our knowledge, is that the presence of strongly
localized models around the bending point has such decisive a
role: the breathers cannot be chosen at any frequency when the
curvature is increased. Their interaction with the linear
localized modes brings about frequencies approximately determined,
and dependent on the curvature.

Consequently, we have developed a variant of the existing
techniques, described in \ref{sec:breathertheory} to obtain
breathers at constant energy. Therefore, the continuation of the
breathers when the curvature is increased, models an adiabatic
process of bending. More biologically meaningful would have been
to model this process at constant temperature, but this is outside
the scope of the techniques used in this paper, which are related
more to single molecule experiments.

The adiabatic process of curving the chain transforms the initial
breather into a nonlinear analogue to the top or bottom linear
mode according to the on--site potential. The corresponding linear
mode, therefore, do not exist in presence of the breather and
cannot produce instabilities. The consequence is that hard
breathers in the bent chain are stable and soft breathers
experience subharmonic instabilities, that conserve, however, the
localization.

We also have shown that up to curvatures relatively high, moving
breathers can travel across the bending point spending some time
at it. They cannot, however be, trapped, which confirms recent
developments in the field \cite{CPAR02b}.

 The physical suggestion is that breathers may be a
means of storage and transport of energy in proteins, due to and
in spite of their complex structure in the space. Certainly, our
model is a rough description of such complex a system, yet the
main conclusion of the present and previous studies could be that
the properties of breathers in different curved chains are quite
common. We also think that they might provide a means to detect
experimentally breathers at the bending points, showing absorbtion
of characteristic frequencies and their harmonics, which depend on
the curvature.  We are, presently, in contact with experimental
groups to explore its feasibility.

\appendix
\section{Details of the model}
\label{sec:Adeduction}


As said in Section~\ref{sec:introduction}, we describe the
position of each amino acid by a vector $\br_n$, $n$ being an
index. The distance between amino acids is supposed to be a
constant $a$, i.e. , $a=|\br_{n+1}-\br_n|$. The direction of the
dipole moments are given by unit vectors %
 $ \bt_n=(\br_{n+1}-\br_n)/|\br_{n+1}-\br_n|$
and the moments themselves are given by $\bp_n=p_n\,\bt_n$.

The interaction energy between two neighbouring dipoles $n$ and $n+1$ is given by
\begin{equation}
\begin{array}{rl}
\label{eq:dipoleU}
 \fl
 U_{n, n+1}
        =&
        \fracc{ \bp_{n+1}\bp_n}{|\br_{n+1}-\br_n|^3}
        -3 \fracc{\bp_n (\br_{n+1} -\br_n) \cdot \bp_{n+1}(\br_{n+1}-\br_n)}{|\br_{n+1}-\br_n|^5}
         \\
\fl
        =&
                p_{n+1} p_n \left( \fracc{\bt_n \bt_{n+1}} {|\br_{n+1}-\br_n|^3}
        -3 \fracc{\bt_n (\br_{n+1} -\br_n) \cdot \bt_{n+1}
        (\br_{n+1}-\br_n)}{|\br_{n+1}-\br_n|^5}\right)
        \\
\fl
        =&
                p_{n+1} p_n \left( \fracc{\bt_n \bt_{n+1}} {a^3}
        -3 a^2 \fracc{\bt_{n} \bt_n \cdot \bt_{n+1}\bt_n}{a^5}\right)  \\
\fl
        =&
        (p_{n+1} p_n/a^3)\left( -2 \bt_{n+1} \bt_{n}\right)=
        (p_{n+1} p_n/a^3)\left(   (\bt_{n+1}-\bt_n)^2  -2    \right)\mbox{}
\end{array}
\end{equation}

Suppose that the equilibrium value of $p_n=p_0,\; \forall n$, then we can
express its non equilibrium value as $p_n=p_0+q\,u_n$, where $u_n$ represent the
stretching from the equilibrium positions.  Substitution into
Equation~\ref{eq:dipoleU}  leads to
\begin{equation}
\label{eq:U012}
 U_{n,n+1}=U^0_{n,n+1}+U^1_{n,n+1}
+\left(-2\fracc{q^2}{a^3}+
\fracc{q^2}{a^3}(\bt_{n+1}-\bt_n)^2\right) u_{n+1} u_{n}\;\;
\end{equation}
The first two terms in this equation stand for constant and linear
terms, therefore, they will not appear in the Hamiltonian, as its first
derivatives with respect to the variables $u_n$ at equilibrium, i.e.,
$u_n=0$ must be zero. We represent the remaining term by $H_{n,n+1}$.

Suppose that the Hamiltonian of the system is given by
\begin{equation}
\label{eq:generichamiltonian}
 H=\sum_n  \fracc{1}{2} {\dot u}^2_n
+\fracc{1}{2}\alpha^2 u_n^2 +\Psi(u_n) + \fracc{1}{2} K
(u_{n+1}-u_{n})^2  + H_{n,n+1} \;,\;
\end{equation}
where $\alpha$ and $K$ are positive constants and $\Psi(u_n)$ is the
nonlinear part of the on--site potential. Rearranging the terms we
obtain
\begin{eqnarray}
\label{eq:rearranged}
 \fl
 \lefteqn{ H=\sum_n \ \fracc{1}{2} {\dot u}^2_n
 +\fracc{1}{2}(\alpha^2-4\fracc{q^2}{a^3})u_n^2 +\Psi(u_n)
\mbox{}}\nonumber
 \\ \lo +\,\fracc{1}{2} (K+2\fracc{q^2}{a^3})
(u_{n+1}-u_{n})^2 + \fracc{q^2}{a^3}(\bt_{n+1}-\bt_n)^2   u_{n+1}
u_{n}
\end{eqnarray}
Thus, the Hamiltonian can be written as
\begin{eqnarray}
\label{eq:Amainhamiltonian}
 \fl
 H=\sum_n \ \fracc{1}{2} {\dot u}^2_n
+\fracc{1}{2}\omega_0^2 u_n^2 +\Psi(u_n) +
  \fracc{1}{2} \ee (u_{n+1}-u_{n})^2  +
\mu (\bt_{n+1}-\bt_n)^2   u_{n+1} u_{n}\;,\nonumber
\end{eqnarray}
with $\omega_0=1$, $\ee=(K+2\,q^2/a^3)/\alpha'^2$, $\mu=q^2/(a^3
\alpha'^2)$, where $\alpha'^2=\alpha^2-4\,q^2/a^3$, with the time
rescaled $t \rightarrow \alpha' t$ and the redefinition of
$\Psi(u_n)\rightarrow\Psi(u_n)/\alpha'^2$. The initial Hamiltonian has
been also divided by $\alpha'^2$,

 We keep the term $w_0$ in the formulae in spite of being $1$ as a
reference because its meaning is the frequency at very low amplitude when the
oscillators are isolated.

The corresponding dynamical equations are
\begin{eqnarray}
\label{eq:Adynamic1}
 \fl
 {\ddot u}_n+
\omega_0^2 u_n +\Psi '(u_n) + \ee (2\,u_n -u_{n+1}-u_{n})
\nonumber\\
 \lo +\, \mu ((\bt_{n+1}-\bt_n)^2 u_{n+1}+ (\bt_{n-1}-\bt_n)^2
u_{n-1})=0,
\end{eqnarray}
which are written in a simplified notation in Equation~\ref{eq:maindynamic}.

\section{Breathers with constant energy}
\label{sec:breathertheory}

To obtain breathers with constant energy, we use a variant of the
method in the time-Fourier space \cite{MA96,M97}. First, we obtain
breathers by path continuation at constant frequency from the
anticontinuous limit using the Newton method as described in the
references above. Thereafter, the path continuation is performed at
constant energy. In this case our variant applies and we give below
the details needed to obtain the corresponding Jacobian used by the
Newton method. It is also a variant of the method developed in
\cite{AMM99} at constant action.

The Hamiltonian of the system can be written:
\begin{equation}
H=\fracc{1}{2}\sum_n \dot{u}_n^2+W(u,\kappa)\,,
\end{equation}
where $u=(u_1,\dots,u_N)$, and $W(u,\kappa)$ represents the potential
energy, sum of the on--site and coupling energies; $\kappa$ represents
any parameter, as the curvature in this article. The dynamical equations
are given by $\ddot{u}_n+\fracc{\partial W}{\partial u_n}=0$.
Time-reversible periodic solutions with frequency $\wb$ are given by a
truncated Fourier series:
\begin{equation}
u_n(t)=\sum_{k=0}^{k_m}(2-\delta_{k,0})\,z_{k,n}\cos(k\,\wb t)
\label{eq:unzk}
\end{equation}
The use of functions with different frequencies is inconvenient and can
be avoided by the change of variable $\tilde{t}=\wb t$, which leads to
the Hamiltonian:
\begin{equation}
H=\fracc{1}{2}\wb^2 \sum_n u_n'^2+W(u,\kappa), \label{eq:hamprime}
\end{equation}
where $\prime{}$ represent the derivative with respect to $\tilde{t}$.
The corresponding dynamical equations are:
\begin{equation}
f_n(u,\wb,\kappa)\equiv \wb^2 u_n''+
\fracc{\partial W}{\partial u_n}=0 \label{eq:dynprime}
\end{equation}
In this equations $\wb$ enters as a parameter and the functions $u_n$,
with frequency unity in the new time variable, are given by:
\begin{equation}
u_n(\tilde{t})=\sum_{k=0}^{k_m}(2-\delta_{k,0})z_{k,n}\cos(k\tilde{t})
\label{eq:unzkprime}
\end{equation}
Therefore, the functions $f_n$ can be seen as $2\pi$-periodic functions
of $\tilde{t}$ due to their dependence on $u(\tilde{t})$ and
$u''(\tilde{t})$.

The components of the cosine discrete Fourier transform of the dynamical
equations (\ref{eq:dynprime}) are given by:
\begin{equation}
F_k[f_n]=-k^2\wb^2 z_{k,n}+F_k [\fracc{\partial W}{\partial
u_n}(u(\tilde{t}))]\,, \label{eq:phikfn}
\end{equation}
where $F_k[g(\tilde{t})]$ represent the k-th term in the cosine
discrete Fourier transform of a $2\pi$ periodic function
$g(\tilde{t})$, given by:
\begin{equation}
F_k[g(\tilde{t})]=\fracc{1}{2\,k_m+1}\sum_{m=0}^{k_m} (2-\delta_{m,0})\,
g(\fracc{2\pi m}{2\,k_m+1})\,
\cos(\fracc{2\pi k\,m}{2\,k_m+1})
\end{equation}
 In this way, finding a breather is reduced to solve the
$(k_m+1)\times N$ equations (\ref{eq:phikfn}). As a breather
solution is determined by the same number of Fourier coefficients
$z_{k,n}$ and the frequency $\wb$, we need and extra equation. If
we are interested in obtaining solutions with given energy $E$,
this is given by $H-E=0$, which for time-reversible solutions
($u_n'(0)=0$) reduces to
 $$
  F_H\equiv W(u(0),\kappa)-E=0
  $$

We denote by $\tilde{z}$ and $\tilde{F}$ the column matrices
$(\{z_{k,n}\},\wb)$ and $(\{F_k[f_n]\},F_H)$, respectively, with the
index $k$ running faster. In order to use the Newton method for path
continuation with the parameter $\kappa$ we need the Jacobian:
\begin{equation}
\fracc{\partial \tilde{F}}{\partial \tilde{z}}=
\left[
\begin{array}{ccc}
\Bigg[\left\{ \fracc{\partial F_k[f_n]}{\partial z_{k',n'}} \right\}
\Bigg]&
  \begin{array}{c} \fracc{\partial F_k[f_n]}{\partial \wb}\\ \vdots
  \end{array}
  \\
\begin{array}{cc}
\left\{\fracc{\partial F_H}{\partial z_{k',n'}}\right\}&
\dots
\end{array}&
\fracc{\partial F_H}{\partial \wb}
\end{array}
   \right]
\label{eq:jacprime}
\end{equation}
The left upper block is the Jacobian at constant frequency,
its elements
given by:
\begin{equation}
 \fl
 \fracc{\partial F_k[f_n]}{\partial z_{k',n'}}=
-k^2\wb^2\delta_{k,k'}\delta_{n,n'}+\fracc{2-\delta_{0,k'}}{2}
\left(F_{k+k'}[\fracc{\partial^2 W}{\partial u_n \partial u_{n'}}]
+F_{|k-k'|}[\fracc{\partial^2 W}{\partial u_n
\partial u_{n'}}]\right)
\end{equation}
The elements in the last column are given by:
\begin{equation}
\fracc{\partial F_k[f_n]}{\partial \wb}=-k^2\wb z_{k,n}\quad;\quad
\fracc{\partial F_H}{\partial \wb}=\fracc{\partial W(u(0))}
{\partial \wb}=0
\end{equation}
The remaining elements in the last row are:
\begin{eqnarray}
 \fl \fracc{\partial F_H}{\partial z_{k',n'}}=
\fracc{\partial W(\{u_n(0)\})}{\partial z_{k',n'}}=
\fracc{\partial W(\{\sum_{l=0}^{k_m}
(2-\delta_{l,0})z_{l,m}\})}{\partial z_{k',n'}}
  \,=\,
 \left(\fracc{\partial W}{\partial u_{n'}}\right)_{t=0}
(2-\delta_{k',0})\,
 \nonumber \\
 \lo=\, -\ddot{u}_{n'}(0)\,(2-\delta_{k',0})\,
 \,=\,(\sum_{l=0}^{k_m} l^2\wb^2
z_{l,n'})\,(2-\delta_{k',0})
\end{eqnarray}

\ack
This work has been supported by the European Commission under the RTN
project LOCNET, HPRN-CT-1999-00163.

JFR Archilla and Yu B Gaididei acknowledge Informatics and Mathematical
Modelling, The Technical University of Denmark, for its hospitality.

\section*{References}
\newcommand{\noopsort}[1]{} \newcommand{\printfirst}[2]{#1}
  \newcommand{\singleletter}[1]{#1} \newcommand{\switchargs}[2]{#2#1}
  \ifx\undefined\allcaps\def\allcaps#1{#1}\fi\ifx\undefined\allcaps\def\allcap%
s#1{#1}\fi

\end{document}